\documentclass[twocolumn,superscriptaddress,showpacs,amsmath,amssymb,aps,pre]{revtex4-1}

\usepackage{graphicx}
\usepackage{dcolumn}
\usepackage{bm}

\usepackage{color}
\usepackage{soul} 

\usepackage{appendix}

\usepackage[pdftex, colorlinks=true, linkcolor=blue, citecolor=blue,
urlcolor=blue]{hyperref}

\usepackage{times}

\graphicspath{{./figs/}}
\DeclareGraphicsExtensions{.pdf}

\begin{document}

\title{Transmission phase of a quantum dot and statistical fluctuations of partial-width amplitudes}

\author{Rodolfo A. Jalabert}\affiliation
{Institut de Physique et Chimie des Mat\'{e}riaux de
Strasbourg, Universit\'{e} de Strasbourg, CNRS UMR 7504, F-67034 Strasbourg, France}
\author{Guillaume Weick}\affiliation
{Institut de Physique et Chimie des Mat\'{e}riaux de
Strasbourg, Universit\'{e} de Strasbourg, CNRS UMR 7504, F-67034 Strasbourg, France}
\author{Hans A. Weidenm\"uller}\affiliation{Max-Planck-Institut f\"ur Kernphysik, D-69029 Heidelberg, Germany}
\author{Dietmar Weinmann}\affiliation
{Institut de Physique et Chimie des Mat\'{e}riaux de
Strasbourg, Universit\'{e} de Strasbourg, CNRS UMR 7504, F-67034 Strasbourg, France}


\begin{abstract}
Experimentally, the phase of the amplitude for electron transmission
through a quantum dot (transmission phase) shows the same pattern
between consecutive resonances. Such universal behavior, found for
long sequences of resonances, is caused by correlations of the signs
of the partial-width amplitudes of the resonances. We investigate the
stability of these correlations in terms of a statistical model. For a
classically chaotic dot, the resonance eigenfunctions are assumed to
be Gaussian distributed. Under this hypothesis, statistical
fluctuations are found to reduce the tendency towards universal phase
evolution. Long sequences of resonances with universal behavior only
persist in the semiclassical limit of very large electron numbers in
the dot and for specific energy intervals. Numerical calculations
qualitatively agree with the statistical model but quantitatively
are closer to universality.
\end{abstract}


\maketitle

\section{Introduction}
\label{Sec:Introduction}

The phase of the transmission amplitude (in short: the transmission
phase) is a key element in the description of coherent transport of
electrons through a quantum dot (QD). The phase is not accessible via
standard conductance measurements
\cite{hackenbroich01,levy95,aharony02}. A breakthrough was achieved
with the advent of phase-sensitive experiments on ballistic
two-dimensional QDs in the Coulomb-blockade regime
\cite{yacoby95}. The QD was placed in one arm of a phase-coherent
ring. The Aharonov-Bohm conductance oscillations measured as a
function of the magnetic flux piercing the ring in an open (or
``leaky") interferometer \cite{schuster97} yielded an indirect
determination of the transmission phase of the QD.

Variation of the plunger gate voltage (and, thereby, of the
electrostatic potential) on the QD made it possible to investigate
sequences of resonances. Long sequences of in-phase resonances were
observed \cite{schuster97} in relatively large QDs (with around 200
electrons on the dot) suggesting {\it universal} behavior. In very
small QDs (with up to 14 electrons) the relative phase of consecutive
resonances appeared to be random \cite{avinum05}. That case was
dubbed the ``mesoscopic regime" (even though both cases
are in the regime of coherent transport which is usually referred to
as the mesoscopic regime).

\begin{figure}[tb]
\centerline{\includegraphics[width=0.68\linewidth]{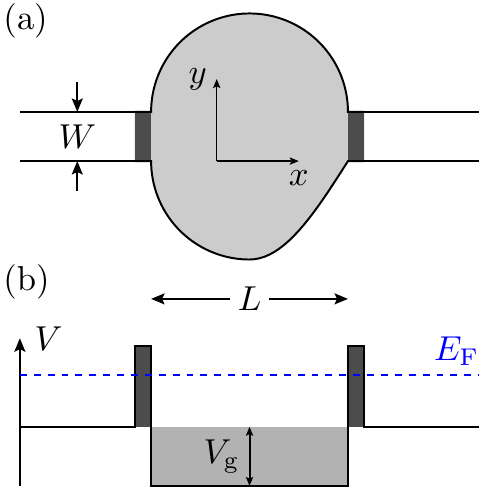}}
\caption{\label{fig:setup} (Color online) (a) Sketch of an asymmetric
  quantum dot (light gray) connected to leads through tunnel barriers
  (dark gray).  (b) Cut of the potential landscape in the
  longitudinal direction. A nearby plunger gate allows to change the
  electrostatic potential $V_\mathrm{g}$ within the dot. The Fermi
  level (blue dashed line) determines the wavenumbers $k_\mathrm{F}$ in the
  leads and $k$ in the dot, through $E_\mathrm{F} = \hbar^2
  k_\mathrm{F}^2 / 2m = \hbar^2 k^2 / 2m - V_\mathrm{g}$.}
\end{figure}

Electron transport through a QD connected to two single-mode leads as
depicted in Fig.~\ref{fig:setup} can be viewed as a quantum scattering
process.  According to the Friedel sum rule, the scattering phase
shift increases by $\pi$ when the electrochemical potential $\mu =
E_\mathrm{F} + V_\mathrm{g}$ is swept through a resonance by changing
the electrostatic potential $V_\mathrm{g}$ on the dot. Here,
$E_\mathrm{F}$ is the Fermi energy in the leads. The transmission
phase follows the scattering phase shift unless the transmission
amplitude has a zero~\cite{lee99,taniguchi99,levy00}.
In that case, the crossing of the origin of the complex plane produces
a phase slip of $\pi$. The experimentally observed phase locking of
resonances in large QDs then necessitates a phase slip of $\pi$ or
equivalently, a zero of the transmission amplitude between every pair
of resonances.  In the literature that situation is indistinctly
referred to as phase locking, phase slip or transmission zero between
consecutive resonances.

The observed phase locking has posed a theoretical puzzle since it
appears to contradict the expectation that eigenstates of different
resonances are uncorrelated. Numerous theoretical papers have
addressed the emergence of universal behavior in large QDs
\cite{hackenbroich01,levy95,aharony02,levy00,
  taniguchi99,lee99,hackenbroich97,baltin99a,baltin99b,silvestrov00,
  kim06,karrasch07a,molina12,molina13}. Some works have pointed to the
importance of electronic correlations in establishing universal
behavior \cite{karrasch07a}, while detailed many-body numerical
calculations recently disputed such a view \cite{molina13}. Other
works described the Coulomb blockade on the QD in terms of the
constant-interaction model, reducing the problem to a single-particle
one \cite{levy00,hackenbroich97,baltin99a,baltin99b,silvestrov00,
  molina12}. The universality of the transmission phase is
then related to the existence of broad levels generated by charging
effects \cite{baltin99b, silvestrov00, kim06} and/or to properties of
the single-particle wavefunctions representing the
resonances~\cite{levy00, hackenbroich97, baltin99a, molina12}. In
particular, it was proposed in Ref.~\cite{molina12} that quantum chaos
on the QD causes spatial correlations of the single-particle
wavefunctions and, thus, is at the root of the experimentally observed
emergence of universality.

In the present paper we critically examine the proposal of
Ref.~\cite{molina12}. Assuming a Gaussian distribution for the
single-particle eigenfunctions of the QD, we calculate the statistical
fluctuations of the lead-dot coupling amplitudes and determine the
probability of zeros of the transmission amplitude. We show that the
fluctuations weaken the tendency towards universality found in
Ref.~\cite{molina12}.

The behavior of the transmission phase is determined by the
partial-width amplitudes (PWAs) of the resonances in the QD. With
consecutive resonances labeled by a running index $n$, the left
(right) PWA of the $n$\textsuperscript{th} resonance with
eigenfunction $\psi_n(x,y)$ reads \cite{alhassid00}
\begin{equation}
\label{eq:gammalr}
\gamma_n^{\mathrm{l}(\mathrm{r})} = \sqrt{\frac{\hbar^2 k_\mathrm{F}
P_\mathrm{c}}{m}}
\int_{0}^{W} {\rm d} y \,\psi_n(x^{\mathrm{l}(\mathrm{r})},y) \, \Phi_{\mathrm{l}(\mathrm{r})}(y)
\ .
\end{equation}
Within the constant-interaction model and under neglect of the
magnetic field in the QD, $\psi_n(x,y)$ can be chosen real. The
geometry is sketched in Fig.~\ref{fig:setup}(a). The leads of width $W$
are connected to the QD by tunnel barriers of transparency
$P_\mathrm{c}$. The distance between entrance and exit leads is $L \gg
W$. The first transversal subband wavefunction in the left
(right) lead is written as $\Phi_{\mathrm{l}(\mathrm{r})}$, and the
integration in Eq.~(\ref{eq:gammalr}) is along the transverse
coordinate $y$ at the entrance (exit) of the QD located at
$x=x^{\mathrm{l}(\mathrm{r})}$. The Fermi wavenumber in the leads is
denoted by $k_\mathrm{F}$, and the effective electron mass by $m$.

We consider the generic case (referred to as \textit{restricted
off-resonance} behavior \cite{molina13}) where the PWAs do not
fluctuate strongly with $n$. Then the behavior of the transmission
amplitude is determined by the PWAs of the two resonances closest in
energy. The transmission amplitude vanishes between the
$n$\textsuperscript{th} and $(n+1)$\textsuperscript{st} resonance if
and only if \cite{levy00}
\begin{equation}
\label{eq:condition}
D_n = \gamma_n^{\mathrm{l}} \gamma_n^{\mathrm{r}}
\gamma_{n+1}^{\mathrm{l}} \gamma_{n+1}^{\mathrm{r}} > 0 \ .
\end{equation}
Then there is an overall phase slip of $\pi$ between the two
resonances. We mention in passing that there are cases where the PWAs
fluctuate strongly (\textit{unrestricted off-resonance} behavior) and
where the criterion (\ref{eq:condition}) does not apply
\cite{molina13}.

\section{Gaussian distribution of partial-width amplitudes}
\label{Sec:Gaussian}

Actual values of the PWAs depend on the geometry of the QD. A generic
description can only be based upon a statistical approach. In this
framework the probability $\mathcal{P}(D_n<0)$ for condition
(\ref{eq:condition}) to be violated has been calculated in various
scenarios (i.e., disordered QDs \cite{levy00} and ballistic chaotic
quantum billiards \cite{molina12}). We follow that line using a
particular statistical hypothesis related to quantum chaos, and we
discuss various parameter regimes.

We define the parity of the $n$\textsuperscript{th} resonance as the
sign of $\gamma_n^{\mathrm{l}} \gamma_n^{\mathrm{r}}$ and the
probability of having a positive parity as
$\mathcal{P}(\gamma_n^{\mathrm{l}} \gamma_n^{\mathrm{r}}>0)$. Under
the assumption that the eigenfunctions of the $n$\textsuperscript{th}
and $(n+1)$\textsuperscript{st} resonances are statistically
uncorrelated, we have
\begin{align}
\label{eq:probDn}
{\cal P}(D_n<0) =&\; 
{\cal P}(\gamma_n^{\mathrm{l}} \gamma_n^{\mathrm{r}} > 0)\big[1 - {\cal P}(\gamma_{n+1}^{\mathrm{l}}
\gamma_{n+1}^{\mathrm{r}} > 0)\big]
\nonumber \\
&+
{\cal P}(\gamma_{n+1}^{\mathrm{l}} \gamma_{n+1}^{\mathrm{r}} > 0)\big[1 - {\cal P}(\gamma_{n}^{\mathrm{l}}
\gamma_{n}^{\mathrm{r}} > 0)\big] \ .
\end{align}
We assume that the classical dynamics of electrons moving
independently in the QD is chaotic. In this case, according to the
Voros-Berry conjecture, the Wigner function is ergodically distributed
on the energy manifold of phase space \cite{voros76,berry77,
  gnutzmann08}.  This assumption implies that the eigenfunction
$\psi_n$ belonging to the eigenvalue $\epsilon_n = \hbar^2 k_n^2 / 2m$
has a Gaussian probability density $p(\psi_n)$ \cite{srednicki96}. For
a two-dimensional billiard with area $\mathcal{A}$ and position vector
$\mathbf{r}=(x,y)$, the probability density is given by
\begin{equation}
\label{eq:probdistrwf}
p(\psi_n) = \mathcal{N} \exp{ \left( -\frac{1}{2} \int_\mathcal{A} \!
\! \mathrm{d} \mathbf{r}
\int_\mathcal{A} \! \! \mathrm{d} \mathbf{r}^\prime \psi_n(\mathbf{r}) 
K(\mathbf{r}, \mathbf{r}^\prime; k_n)\psi_n (\mathbf{r}^\prime)
\right)}
\end{equation}
where $\mathcal{N}$ is the normalization constant. The function $K$ is
defined by
\begin{equation}
\label{eq:defK}
\int_\mathcal{A} \mathrm{d} \mathbf{r} \ K(\mathbf{r}, \mathbf{r}^\prime;
k) \ J_0(k|\mathbf{r} - \mathbf{r}^\prime|) = \mathcal{A} \ 
\delta(\mathbf{r}-\mathbf{r}^\prime) \ ,
\end{equation}
where $J_0$ is the zeroth Bessel function of the first kind. Equation
\eqref{eq:probdistrwf} implies $\langle \psi_n \rangle = 0$ and a
correlation of the values of the eigenfunction $\psi_n$ at two points
$\mathbf{r}$ and $\mathbf{r}^\prime$ given by \cite{berry77}
\begin{equation}
\label{eq:berry}
\langle \psi_n(\mathbf{r}) \psi_n(\mathbf{r}^\prime) \rangle =
\frac{1}{\mathcal{A}} \ J_0(k_n|\mathbf{r}-\mathbf{r}^\prime|) \  .
\end{equation}
The angular brackets denote the average over $p(\psi_n)$. The
eigenfunctions belonging to different resonances are uncorrelated, so
that $p(\psi_1, \psi_2, \ldots) = \prod_n p(\psi_n)$.

We recall that the spatial correlation of wavefunctions has also been
derived from information theory \cite{jarzynski97} or, in the case of
weakly disordered systems, with the aid of supersymmetry techniques
\cite{prigodin95,mirlin}. The effects of spectral, position, and
directional averages in expression \eqref{eq:berry} have been
thoroughly discussed in Refs.~\cite{toscano} and \cite{li2001}.
Furthermore, this important relation has been experimentally tested in
the eigenmodes of resonant microwave cavities \cite{stockmann} and
numerically checked in different dynamical systems
\cite{steiner93,li94,basch98,basch2002}, especially in the context of
the so-called rate of quantum ergodicity (i.e., the rate in which the
quantum-mechanical expectation value tends to its mean value upon
approaching the semiclassical limit of large energies). Along these
lines, Srednicki and Stiernelof~\cite{srednicki96b} used the Gaussian
hypothesis \eqref{eq:probdistrwf} to show that the root-mean-square
amplitude of the statistical fluctuations around the mean value given
by Eq.~\eqref{eq:berry} decrease in the semiclassical limit as $(k^2
\mathcal{A}_{\mathrm{R}})^{-1/4}$, where $\mathcal{A}_{\mathrm{R}}$ is
the area of the billiard used for a spatial average of the
autocorrelator ($\mathcal{A}_{\mathrm{R}} \ll \mathcal{A}$).

According to Eqs.~\eqref{eq:gammalr} and \eqref{eq:probdistrwf},
each PWA is the sum of Gaussian-distributed amplitudes and, hence, has
a Gaussian distribution, too. From $\langle \psi_n \rangle = 0$ and
$\langle \psi_n \psi_{n'} \rangle$ $= 0$ for $n \neq n'$ we have
$\langle \gamma_n^{\mathrm{l}(\mathrm{r})} \rangle = 0$ and $\langle
\gamma_n^{\mathrm{l}(\mathrm{r})} \gamma_{n'}^{\mathrm{l}(\mathrm{r})}
\rangle = 0$ for $n \neq n'$. For each $n$ the distribution of the
PWAs is then characterized by the three second moments $\langle
\gamma_n^{\mathrm{l}} \gamma_n^{\mathrm{l}} \rangle$, $\langle
\gamma_n^{\mathrm{r}} \gamma_n^{\mathrm{r}} \rangle$, and $\langle
\gamma_n^{\mathrm{l}} \gamma_n^{\mathrm{r}} \rangle$. Left-right symmetry 
of the couplings between the leads and the QD implies the equality
\begin{equation} 
\label{eq:sigma_def}
\sigma_n^2 = \langle \gamma_n^{\mathrm{l}}\gamma_n^{\mathrm{l}} \rangle =
\langle \gamma_n^{\mathrm{r}}\gamma_n^{\mathrm{r}} \rangle \ .
\end{equation}
With
\begin{equation} \label{eq:rho}
\rho_n = \frac{1}{\sigma_n^2} \
\langle \gamma_n^{\mathrm{l}}\gamma_n^{\mathrm{r}} \rangle \ ,
\end{equation}
the joint probability density of the left and right PWA is
\begin{align}
\label{eq:jointprobdistr}
p(\gamma_n^{\mathrm{l}}, \gamma_n^{\mathrm{r}}) =&\,
\frac{1}{2 \pi \sigma_n^2 \sqrt{1 - \rho_n^2}} 
 \nonumber \\
&\times
 \exp{\left( - \frac{\left(\gamma_n^{\mathrm{l}}\right)^2+
\left(\gamma_n^{\mathrm{r}}\right)^2 - 2 \rho_n \gamma_n^{\mathrm{l}}
\gamma_n^{\mathrm{r}}}{2 \sigma_n^2 (1 - \rho_n^2)} \right)} \ .
\end{align}
The probability for the product $\gamma_n^{\mathrm{l}} \gamma_n^{\mathrm{r}}$ to
be positive is obtained from Eq.~\eqref{eq:jointprobdistr} as
\begin{equation}
\label{eq:pospar}
\mathcal{P}(\gamma_n^{\mathrm{l}} \gamma_n^{\mathrm{r}} > 0) =
\frac{1}{2} + \frac{1}{\pi} \arcsin{(\rho_n)} \ .
\end{equation} 
Completely
correlated (anti-correlated) PWAs corresponding to $\rho_n = 1$ ($ -1
$) lead to $\mathcal{P}(\gamma_n^{\mathrm{l}} \gamma_n^{\mathrm{r}} >
0) = 1$ ($0$, respectively), while in the uncorrelated case we have
$\rho_n = 0$ and $\mathcal{P}(\gamma_n^{\mathrm{l}}
\gamma_n^{\mathrm{r}} > 0) = 1/2$.

For the evaluation of Eq.~(\ref{eq:probDn}) we have to determine the
dependence of $\rho_n$ on $k_n$. With the QD being chaotic, the
distribution of spacings $\epsilon_n - \epsilon_{n + 1}$ of nearest
eigenvalues is given by the Wigner surmise. However,
$\rho_n$ is expected to be a smooth function of $k_n$ on the scale of
the mean wavenumber difference $\Delta k_n = \pi/k_nL^2$. Therefore,
\begin{equation}
\label{eq:probDn2}
{\cal P}(D_n<0) \simeq 2 f(k_n) + \Delta k_n f'(k_n)
\end{equation} 
with
\begin{equation}
\label{eq:functf}
f(k) =
\frac{1}{4}-\frac{1}{\pi^2}\arcsin^2{\big(\rho(k)\big)}\ .
\end{equation}
The extreme values of $\rho$ are $\rho = \pm 1$. Therefore,
$\mathrm{d} \rho/\mathrm{d} k = 0$ for $|\rho| = 1$, and the
expression~(\ref{eq:probDn2}) is well defined for all values of $\rho$.

Equation \eqref{eq:probDn2} shows that there are two possible reasons
for violations of the universal behavior ${\cal P}(D_n < 0) = 0$. (i)
The condition $|\rho| = 1$ may be violated so that
$\gamma_n^{\mathrm{l}}$ and $\gamma_n^{\mathrm{r}}$ are not perfectly
correlated or anticorrelated, and $f(k_n) \ne 0$. (ii) Even if the
previous condition is met, $\Delta k_n$ may not be negligible. Reason
(i) becomes the dominant one in the semiclassical regime
\cite{molina12}, where $\Delta k_n \propto 1/k_n L^2$, or when the
spectral average over the resonances is taken. We return to that point
in Sec.~\ref{Sec:evsa}.

\begin{figure*}
\includegraphics[width=.9\linewidth]{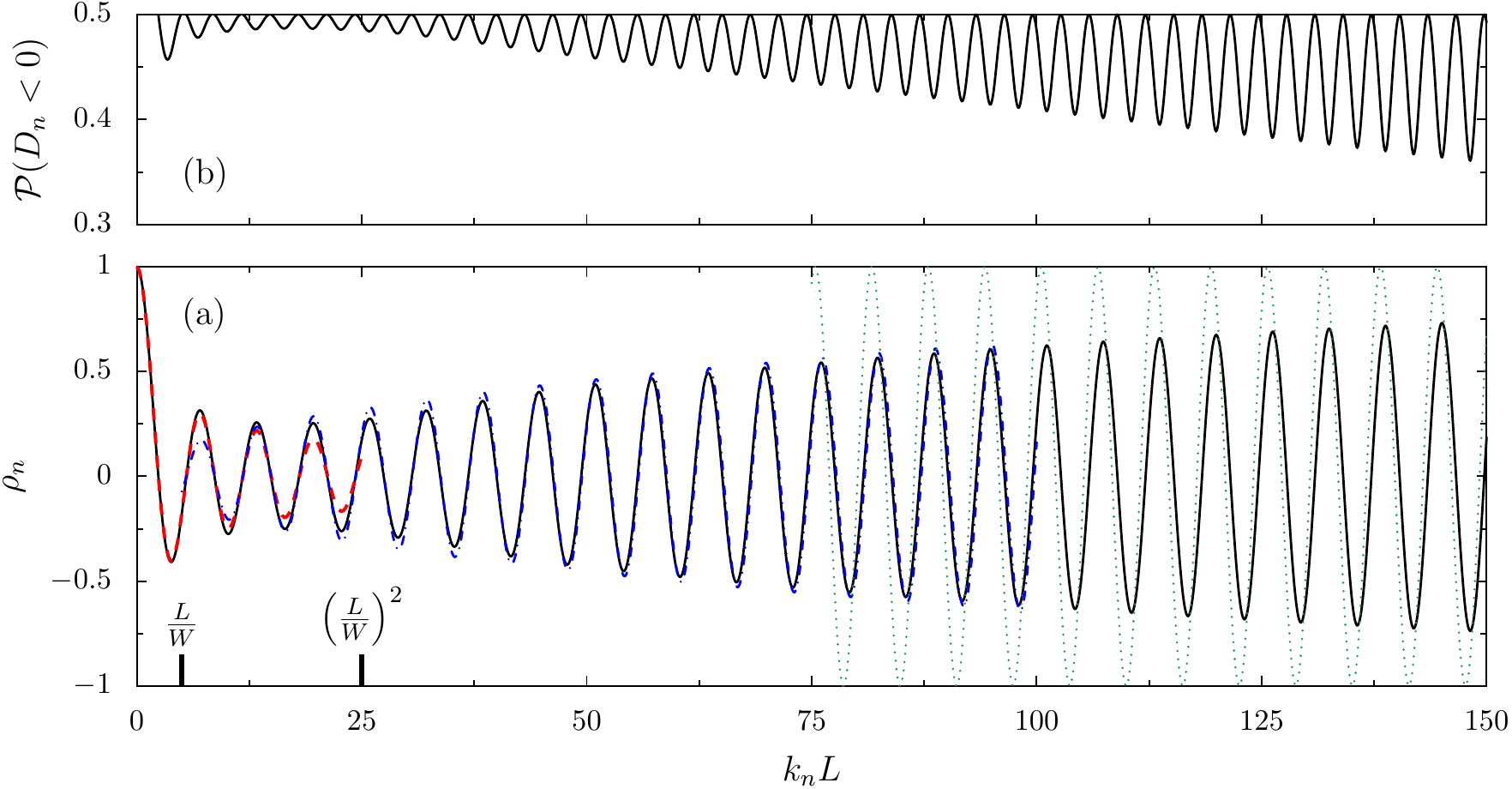}
\caption{\label{fig:rhoandp} (Color online) (a) The
  correlator $\rho_n$ defined in Eq.\ \eqref{eq:rho}, calculated
  numerically from the integrals in Eqs.~\eqref{eq:sigmaint} and
  \eqref{eq:prodgamint}, is plotted as a function of $k_n L$ for $L /
  W = 5$ (black solid line).  The dashed red line represents the
  approximate result of Eq.~\eqref{eq:rho_one} valid for $k_n L
  \lesssim L / W$. The dash-dotted blue line represents the result of
  Eq.~\eqref{eq:rho_few} for the intermediate regime, and the green
  dotted line gives the semiclassical result of
  Eq.~\eqref{eq:rho_sc}. (b) The probability $\mathcal{P}(D_n
  < 0)$ as a function of $k_n L$, calculated from
  Eqs.~\eqref{eq:probDn2} and \eqref{eq:functf} and the numerical
  result for $\rho_n$.}
\end{figure*}

\section{Second moments of partial-width amplitudes}
\label{Sec:cpwd}

We use Eqs.~\eqref{eq:gammalr} and (\ref{eq:berry}) to calculate
$\sigma_n^2$ and $\rho_n$ as defined in Eqs.~(\ref{eq:sigma_def}) and
(\ref{eq:rho}). We assume that QD and leads have hard walls. The first
transversal subband wavefunction for the left (right) lead then reads
\begin{equation}
\Phi_{\mathrm{l}(\mathrm{r})}(y) = \sqrt{\frac{2}{W}} \ 
\sin{\left(\frac{\pi y}{W}\right)} \ ,
\end{equation}
and we have
\begin{align}
\label{eq:sigmaint}
\sigma_n^2 =&\; \alpha \int_0^W\!\!\mathrm{d}y \int_0^W\!\!\mathrm{d}y' \,
\langle \psi_n(x^{\mathrm{l}},y) \psi_n(x^{\mathrm{l}},y') \rangle 
\nonumber \\
&\times\left[\cos{\left(\frac{\pi}{W}(y'-y)\right)} 
-\cos{\left(\frac{\pi}{W}(y'+y)\right)} \right] 
\nonumber \\
=&\; 
\frac{2\alpha W^2}{\mathcal{A}} \int_0^1\!\! \mathrm{d}z \,
J_0(k_n W z)
\nonumber\\
&\times\left[(1-z)\cos{\left(\pi z\right)} + \frac{1}{\pi}\sin{\left(\pi
z\right)} \right] 
 \ ,
\end{align}
where $\alpha = \hbar^2 k_\mathrm{F} P_\mathrm{c} /m W$. We have
introduced dimensionless integration variables, changed to their
difference $z$ and half their sum, and integrated over the latter
variable. For $k_n W \ll 1$ we approximate the argument of $J_0$ in
Eq.\ \eqref{eq:sigmaint} by unity, obtaining
\begin{equation}
\label{eq:sigma_l}
\sigma_n^2 \simeq \frac{\alpha W^2}{\mathcal{A}} \ \frac{8}{\pi^2} \ ,
\end{equation}
while for $k_n W \gg 1$ the integral over $z$ is strongly suppressed
because of the oscillating character of $J_0$. We show in the Appendix
that
\begin{equation}
\label{eq:sigma_sc}
\sigma_n^2 \simeq \frac{\alpha W^2}{\mathcal{A}} \ \frac{2}{k_n W} \ .
\end{equation}
For the correlator $\langle \gamma_n^{\mathrm{l}} \gamma_n^{\mathrm{r}}
\rangle$ we obtain analogously
\begin{align}
\label{eq:prodgamint}
\langle \gamma_n^{\mathrm{l}} \gamma_n^{\mathrm{r}} \rangle =&\; 
\frac{2 \alpha W^2}{\mathcal{A}} \int_{0}^{1} \mathrm{d} z 
\,J_0 \left(k_n L \sqrt{1 + (W/L)^2 z^2}\right)
\nonumber\\
&\times
\left[(1 - z) \cos{\left(\pi z \right)} + \frac{1}{\pi} \sin{\left(
\pi z \right)} \right]
\ .
\end{align}
Analytical results for $\langle \gamma_n^{\mathrm{l}}
\gamma_n^{\mathrm{r}} \rangle$ are obtained in the following
regimes. For $k_n W \ll L / W$ we have
\begin{equation}
\label{eq:prodgam_l}
\langle \gamma_n^{\mathrm{l}} \gamma_n^{\mathrm{r}} \rangle \simeq
\frac{\alpha W^2}{\mathcal{A}} \ \frac{8}{\pi^2} \ J_0(k_n L)\ ,
\end{equation}
while for $k_n W \gg L / W \gg 1$ we show in the Appendix that
\begin{equation}
\label{eq:prodgam_sc}
\langle \gamma_n^{\mathrm{l}} \gamma_n^{\mathrm{r}} \rangle \simeq
\frac{\alpha W^2}{\mathcal{A}} \ \frac{2}{k_n W} \ \cos{(k_nL)} \ .
\end{equation}
The corresponding results for the correlator $\rho$ are obtained by
combining the results~(\ref{eq:sigma_l}) and (\ref{eq:sigma_sc}) with
Eqs.~(\ref{eq:prodgam_l}) and (\ref{eq:prodgam_sc}). The value of
$k_n$ defines three regimes that are depicted in
Fig.\ \ref{fig:rhoandp}(a) for the case $L/W=5$. These are
\begin{itemize}
\item[(i)] the one-mode regime $1< k_nL < L / W$ (red dashed line),
where 
\begin{equation}\label{eq:rho_one}
\rho_n \simeq J_0(k_n L)\ ;
\end{equation}
\item[(ii)] the intermediate regime $L / W < k_nL < (L / W)^2$ (blue
dash-dotted line), where 
\begin{equation}\label{eq:rho_few}
\rho_n \simeq \frac{4}{\pi^2} \ k_n W \ J_0(k_n L)\ ;
\end{equation}
\item[(iii)] the semiclassical regime $k_n L\gg (L / W)^2$ (green dotted
line), where 
\begin{equation}\label{eq:rho_sc}
\rho_n \simeq \cos{(k_n L)} \ .
\end{equation}
\end{itemize}
In addition, the value of $\rho_n$ obtained by numerical evaluation of
Eqs.~\eqref{eq:sigmaint} and \eqref{eq:prodgamint} is shown as a black
line in Fig.\ \ref{fig:rhoandp}(a). The oscillation of
$\rho_n$ around zero is due to the Bessel function in the integrand of
Eq.~\eqref{eq:prodgamint}. The amplitude approaches unity in the
semiclassical regime ($k_n L \gg 1$). 
Figure \ref{fig:rhoandp}(b) shows the resulting probability
$\mathcal{P}(D_n < 0)$ calculated from $\rho_n$ using
Eqs.~\eqref{eq:probDn2} and \eqref{eq:functf}. For the parameters
chosen, the contribution of the second term on the right-hand side of
Eq.~\eqref{eq:probDn2} is significant only for $k_n L \lesssim 10$.
Therefore, the deviation of ${\cal P}(D_n<0)$ from zero is almost
exclusively due to the lack of perfect correlation between the PWAs
belonging to neighboring resonances.

We recall that the condition for adjacent resonances to cause with
probability 1 a lapse in the phase of the transmission amplitude is
given by ${\cal P}(D_n < 0) = 0$. According to
Eqs.\ \eqref{eq:probDn2} and \eqref{eq:rho_sc} this condition is met
only in the semiclassical limit for distinct values of $k$ for which
\begin{equation}
\label{eq:P_sc}
\mathcal{P}(D_n<0)\simeq
\frac 12-2\left(\left\{\frac{k_nL}{\pi}\right\}-\frac 12\right)^2 \ .
\end{equation}
Here $\{x\}$ denotes the fractional part of $x$. Extrapolation of the
data shown in Fig.\ \ref{fig:rhoandp}(b) to larger
values of $k_nL$ suggests that $k$ intervals which meet that condition
do indeed exist.

\section{Ensemble average versus spectral average --- numerical results}
\label{Sec:evsa}

The results in Sec.~\ref{Sec:cpwd} and in Fig.~\ref{fig:rhoandp}
represent averages over the Gaussian ensemble defined in
Sec.~\ref{Sec:Gaussian}. How are we to relate these averages with
actual data obtained by measurements of a single QD (and not on an
ensemble of QDs)? The answer would be simple if ${\cal P}(D_n < 0)$
were independent of $k$ as we could then employ the usual ergodicity
argument and equate the ensemble average obtained in the statistical
approach with the running average of data over $k$. However, the
oscillations of ${\cal P}(D_n < 0)$ away from the completely
uncorrelated value $1 / 2$ towards smaller values shown in 
Fig.~\ref{fig:rhoandp}(b) increase as we approach the
semiclassical limit $k_n L \rightarrow \infty$. Therefore, the actual
value of ${\cal P}(D_n<0)$ becomes increasingly dependent on $k$, and
the relation between the two averages acquires crucial importance.

First, we may think of the correlator in Eq.~(\ref{eq:berry}) as being
the result of an averaging process performed on the actual
eigenfunction of the $n^{\rm th}$ resonance for fixed distance
$|\mathbf{r} - \mathbf{r}^\prime|$. Such spatial averaging, when
performed over a domain larger than the de Broglie wavelength,
improves the rate of quantum ergodicity \cite{srednicki96b,basch2002}
by suppressing the fluctuations around the mean value of the
wavefunction product under consideration. The integrals over $y$ and
$y'$ in the defining Eq.~(\ref{eq:sigmaint}) for $\sigma^2_n$ and in
the corresponding expression for $\langle \gamma_n^{\mathrm{l}}
\gamma_n^{\mathrm{r}} \rangle$ partly amount to such an average.

This argument is purely \textit{ad hoc}, however. Moreover, it does not resolve
the issue of the dependence of ${\cal P}(D_n < 0)$ on $k$. The actual
value of $k$ in the experiments is not known. To make up for that, an
average of ${\cal P}(D_n < 0)$ over one period in $k_n L$ was
considered in Ref.~\cite{molina12}. Under that proposal, the second
term on the right-hand side of Eq.~\eqref{eq:probDn2} yields in the
semiclassical limit a negligible contribution since $f$ becomes a
periodic function of $k$. Using Eq.\ \eqref{eq:P_sc}, the first term
yields $1 / 3$ on average, rendering the occurrence of long sequences
of in-phase resonances quite unlikely. However, averaging over an
entire period in $k_n L$ is not necessary. Indeed, equality of
ensemble average and running average is guaranteed provided the latter
extends over a sufficiently large set of resonances. The average
spacing $\Delta k_n = \pi / k_n L^2$ of resonances becoming very small
in the semiclassical limit, it suffices to consider an averaging
interval much smaller than a full period in $k_n L$ to obtain a
meaningful average. Since the last term on the right-hand side of
Eq.~\eqref{eq:probDn2} is semiclassically negligible, long sequences
of in-phase resonances do exist for $k$ values where $|\rho|$ is close
to unity and ${\cal P}(D_n<0)$ is close to zero. The length of such
sequences of in-phase resonances decreases as ${\cal P}(D_n < 0)$
deviates from zero.

In Ref.~\cite{molina12}, numerical calculations done for the
configuration of Fig.\ \ref{fig:setup} versus plunger gate voltage
yielded for large values of $kL$ long sequences of in-phase
resonances. The difference between the number of resonances and the
number of transmission zeros in a given $k$-interval was found to
become progressively small in the semiclassical limit. Here we report
on a more systematic numerical study. We calculate the distribution of
$D_n$ [cf.\ Eq.\ \eqref{eq:condition}] and compare that with the
probability ${\cal P}(D_n < 0)$ predicted by the Gaussian hypothesis
[Eqs.~\eqref{eq:probDn2} and \eqref{eq:functf}].

\begin{figure}
\includegraphics[width=\linewidth]{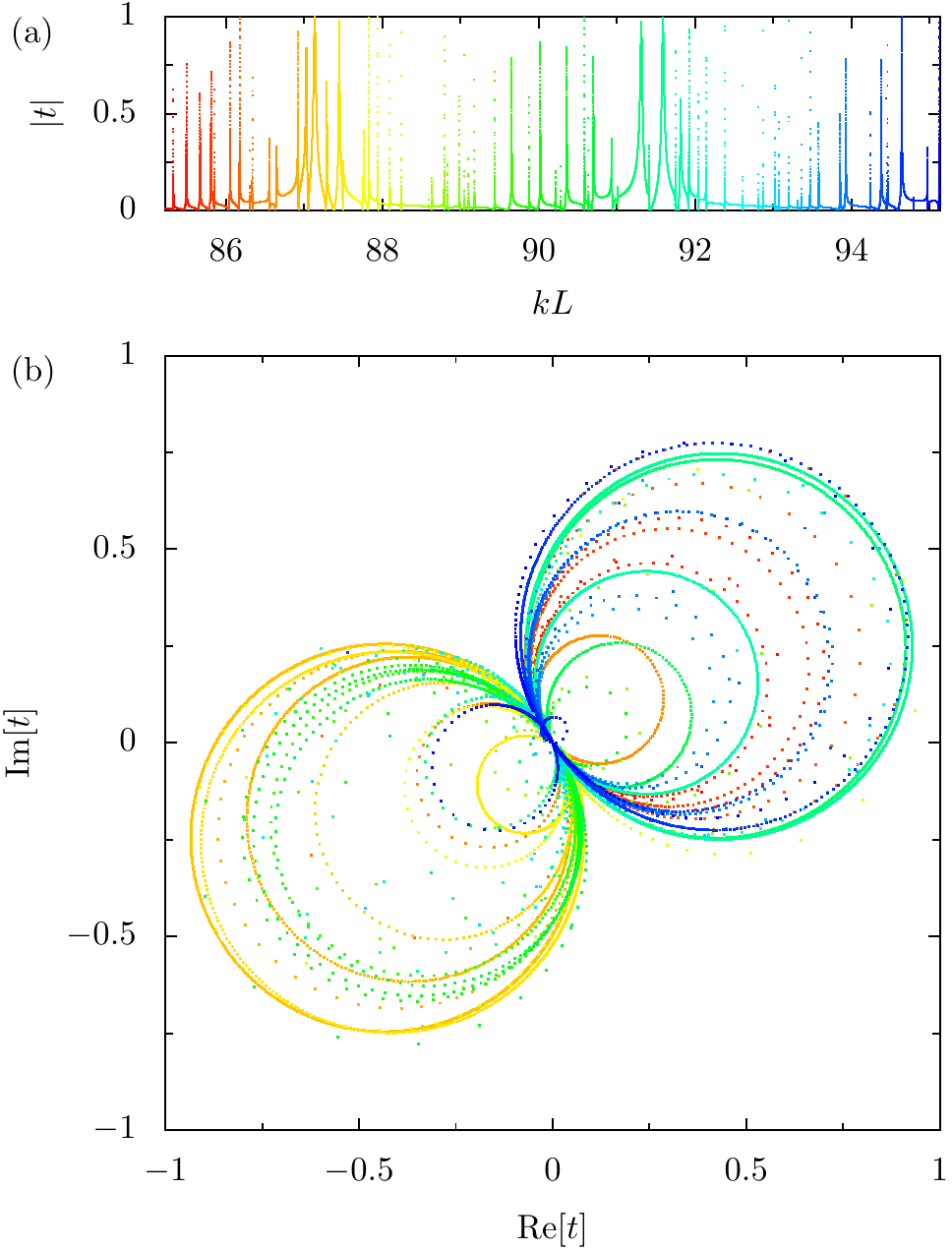}
\caption{\label{fig:t} (Color online) (a) Absolute value of the transmission 
  amplitude $t$ for the setup in Fig.~\ref{fig:setup}, as a function of $kL$. 
  (b) The transmission amplitude presented in the complex plane, for the same 
  values of $kL$ (which are encoded by the color of the data points).
}
\end{figure}
When the plunger gate voltage $V_\mathrm{g}$ is varied over a
sufficiently large interval, the $k$ dependence of the complex
transmission amplitude $t$ displays a sequence of peaks.
Figure \ref{fig:t} shows an example of such a sequence, whose length 
corresponds to changing $kL$ by about $3\pi$. Figure \ref{fig:t}(a) presents 
the peaks of the absolute value of $t(k)$, and Fig.\ \ref{fig:t}(b) shows that $t$ 
approximately follows circles in the complex plane, indicating that most of the 
peaks have Breit-Wigner form. 
At the $n$\textsuperscript{th} resonance we accordingly use
\begin{equation}
\label{eq:Breit-Wigner}
t(k) = \sum_{n} \frac{\gamma_n^{\mathrm{l}} \gamma_n^{\mathrm{r}}}{\epsilon(k)
- \epsilon_n + i \Gamma_n/2}
\end{equation}
with $\Gamma_n = |\gamma_n^{\mathrm{l}}|^2+|\gamma_n^{\mathrm{r}}|^2$
to extract the product $\gamma_n^{\mathrm{l}} \gamma_n^{\mathrm{r}}$
and to obtain $D_n$.

In the complex plane of Fig.\ \ref{fig:t}(b) we can easily recognize the 
relatively broad peaks, represented by a dense set of points along a 
circle, representing data for increasing values of $kL$, while the very 
sharp ones correspond to only a few points on the chosen $kL$ grid. 
When there is a transmission zero between two peaks, $t$ continues to 
turn anticlockwise in the same half-plane. If there is a finite minimal 
value of $|t|$ between two peaks, a switch of half-plane occurs before 
turning (also counter-clockwise) for the $kL$ values corresponding to 
the second peak. The tendency towards universality is already noticeable 
in this restricted sequence of peaks for a single stadium. Peaks of 
similar color (close in $k_nL$) tend to stay in one of the half-planes, 
but there are occasional switches between the two half-planes.  

To determine the distribution of $D_n$ with sufficiently good
statistics, we have taken two steps. First, we have combined a
sequence of $D_n$ values within some $k_n$ interval much 
larger than the level spacing. 
Second, we have combined data generated from stadia with different 
shapes. We describe these steps in turn. For fixed stadium shape, 
the length $\delta / L$ of the sampling interval is bounded from 
above by the fact that according to Fig.~\ref{fig:rhoandp}(b)
the distribution of $D_n$ is expected to be an oscillatory function 
of $k$. 
Values of $\delta \approx \pi$ would mix different distribution 
patterns. On the other hand, values of $\delta \ll \pi$ reduce the 
number of resonances in the interval and increase the statistical error. 
We have chosen $\delta = \pi/4$. We have improved the local statistics 
by combining data from six successive $k_n$ intervals given by 
$[k_n+j\pi/L, k_n+j\pi/L +\delta]$ with $j={-3,-2,-1,0,1,2}$. Since the 
smooth oscillations of $\mathcal{P}(D_n < 0)$ are expected to be
quasi-periodic with period $\pi/L$, that procedure should not cause
any problems. We have used that procedure for a total of fifteen different
stadium shapes. Fourteen of these were generated from Fig.~\ref{fig:setup}(a)
by deforming different quarter circles of the stadium. Each shape
yielded a sequence of peaks that was uncorrelated from the
others~\cite{footnote:shift}.

\begin{figure}
\includegraphics[width=\linewidth]{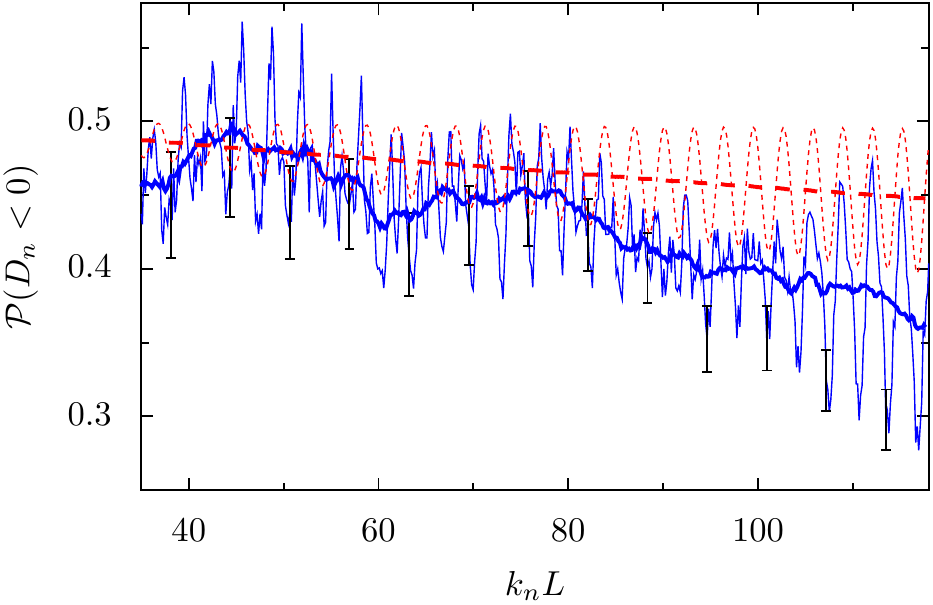}
\caption{\label{fig:numerics} (Color online) Blue solid lines:
  numerically obtained $\mathcal{P}(D_n < 0)$ (see text) using a
  smoothing $k_n$ interval of $\delta/L$ with $\delta=\pi/4$ and $\pi$ 
  (thin and thick lines, respectively). 
  The error bars of some of the data points for
  the smaller smoothing interval indicate the statistical error. 
  Red dashed lines: $\mathcal{P}(D_n < 0)$ from the statistical model,
  Eqs.~\eqref{eq:probDn2} and \eqref{eq:functf}, using smoothing
  intervals with $\delta=\pi/4$ and $\pi$ (thin and thick lines,
  respectively).}
\end{figure}

Figure \ref{fig:numerics} shows $\mathcal{P}(D_n < 0)$ obtained
numerically as described above (thin blue solid line),
together with the analytic result of Eqs.\ \eqref{eq:probDn2} and
\eqref{eq:functf}, locally averaged over $k_n$ intervals of length 
$\delta/L$ with $\delta = \pi/4$ (thin red dashed line). 
The thin red dashed line is a smoothed version of the solid line in  
Fig.\ \ref{fig:rhoandp}(b). The difference between the two 
curves is very small. We have assigned statistical error bars to some 
points of the numerically generated $\mathcal{P}(D_n < 0)$. 
Consistently with our previous analysis we find for the lowest part 
of the interval shown that the statistical errors are quite significant, 
while for larger values of $k_n$ we can be confident that the features 
observed in the curve $\mathcal{P}(D_n < 0)$ are statistically robust. 
For instance, the $\pi/L$ quasi-periodicity of $\mathcal{P}(D_n < 0)$ 
is clearly seen in the thin blue solid curve of Fig.\ \ref{fig:numerics}.

Figure \ref{fig:numerics} shows that for $k_nL \gtrsim 60$ the
numerically generated values of $\mathcal{P}(D_n < 0)$ are
systematically smaller than those of the statistical model. This is
displayed very clearly by the two thick solid blue and dashed red lines obtained by
averaging the data in $k_n$-intervals of length $\pi/L$. We recall
that the statistical model yields a saturation value of $1/3$ only in
the semiclassical limit while the numerical results attain that value
already for $k_nL \simeq 120$. The early approach of the numerical
results to the universality condition $\mathcal{P}(D_n < 0) = 0$
predicted semiclassically may have important consequences in the
analysis of the experiments and for wavefunction correlations.

The surprising quantitative discrepancy between our numerical results 
and the predictions of the statistical model may be due to several 
assumptions of the latter model that may not fully apply. Firstly, 
there is a correction to Eq.~\eqref{eq:berry} that is due to classical
trajectories~\cite{srednicki98}. This correction is significant when
the two arguments of the wavefunction correlator are widely separated.
In principle the correction could be taken into account within a
semiclassical numerical approach. Secondly, Eq.~\eqref{eq:berry} does
not account for boundary effects. The boundary conditions for the
resonance wavefunctions are of mixed Dirichlet-Neumann type (vanishing
$\psi_n$ along the hard walls of the billiard and vanishing normal
derivative at the points connecting the dot to the leads). The ensuing
corrections could in principle be calculated following the
prescription of Ref.~\cite{urbina04}. Thirdly, the Gaussian hypothesis
is perhaps not appropriate to describe the higher moments of the
wavefunction distribution, since it is based on an ergodic
distribution on the energy shell. Such an assumption has its
limitations since it has been shown that, in a two-dimensional
phase-space, a delta-function on the energy shell cannot represent a
positive operator in the Weyl representation \cite{balazs}. That is, a
delta-function cannot be a true Wigner function. This result has been
recently generalized~\cite{ozorio} to any curved surface of dimension
$2d-1$ in a phase-space of dimension $2d$. We have not investigated
any of these challenging issues yet.

\section{Conclusions}
\label{Sec:conclusions}

Leaky Aharonov-Bohm interferometers have given access to the
transmission phase of a quantum dot placed in one of the arms of the
interferometer. Long sequences of in-phase resonances observed
experimentally pose a theoretical puzzle. Assuming that the Coulomb
blockade in the dot can be treated within the constant-interaction
model, we have reduced the theoretical calculation of the transmission
phase to a one-body problem. The phase evolution between neighboring
resonances is determined by their parities. For each resonance the
parity is defined as the sign of the product of the partial-width
amplitudes for entrance and exit leads and is determined directly by
the resonance eigenfunction. This chain of thought has led us from a
flagship problem in mesoscopic physics to one of the fundamental
issues of quantum chaos, that is, the statistical properties of
wavefunctions of a system which is classically chaotic.

Assuming a Gaussian distribution for the eigenfunctions of a
classically chaotic dot, we have calculated the probability of having
in-phase resonances. We have done so for different regimes defined by
the ratio of the de Broglie wavelength $1/k$ in the dot and the length
scales of the problem (the width $W$ of the leads and the distance $L$
between the entrance and exit leads). We found that the fluctuations
of the partial width amplitudes are relevant. Complete in-phase
behavior of the resonances is obtained only in the semiclassical limit
of large $k L$. Even in this regime, in-phase behavior is not
universal but occurs only within certain energy intervals. Numerical
calculations yield qualitatively the same behavior. However, with
increasing $k L$ the numerical results tend to be systematically
smaller than predicted by the statistical model and, thus, closer to
universal behavior.

We stress that the present attempt to explain the experimental results
of Refs.~\cite{yacoby95,schuster97} within the statistical model or
our numerical calculations is in line with much work in mesoscopic
physics. The constant-interaction model is used to reduce the physics
to that of a single-particle problem. The statistical model further
assumes that the chaotic nature of the single-particle classical
dynamics in the quantum dot justifies the use of the Voros-Berry
conjecture and random-matrix theory. It is important to recall that
such an approach has led to a successful description of the
statistical distribution of the height of the Coulomb blockade peaks
\cite{jalabert92,chang96,folks96}. Moreover, the long-range (in
energy) modulation of the peak-height distribution found in some of
the experiments (not accounted for in the simplest random-matrix
description) is found \cite{narimanov01} to be due to spatial
correlations of the resonance wavefunctions described by
Eq.~\eqref{eq:berry}.

It is tempting to think of improvements of the numerical calculations
that might lead to a better agreement with the experimental data. A more realistic
model of the quantum dot could be helpful in yielding information
concerning the regime in which the dot
operates (see Sec.\ \ref{Sec:cpwd}). Such calculations would encounter a number of difficulties,
however. Neither the precise form of the self-consistent
single-particle confinement potential of the dot nor its modification
due to a change of the plunger voltage are known. It seems likely that
the actual situation is quite different from that sketched in
Fig.~\ref{fig:setup} where a change of $V_\mathrm{g}$ merely shifts
the floor of the potential. In fact, a deformation of the confinement
potential caused by changing the plunger voltage has been held
responsible for the phase locking of consecutive resonances
\cite{hackenbroich97} and for the energy modulation of the peak-height
distribution~\cite{vallejos99}. In work using density-functional
theory to calculate the electronic structure of lithographically
defined quantum dots~\cite{stopa96} it was found that some properties
(like the peak-height distribution) are relatively robust with respect
to details of the confining potential, while others (like the energy
modulation of the peak heights) are not. Still we believe that more
realistic models for quantum dots, together with the study of
wavefunction fluctuations beyond the Gaussian assumption, appear as
promising avenues opened by the present investigation.

\acknowledgments

We thank A.\ M.\ Ozorio de Almeida for helpful discussions and 
for letting us know unpublished results \cite{ozorio}, and R.\ A.\ Molina for useful remarks on the
manuscript. We acknowledge the financial support from the ANR through grant ANR-08-BLAN-0030-02.

\appendix
\section*{Appendix: Approximate integrations in the semiclassical limit}
\label{Sec:appendix}
We evaluate Eqs.\ \eqref{eq:sigmaint} and \eqref{eq:prodgamint} in the
semiclassical limit. Introducing the variable $x=k_nWz$, we write
Eq.\ \eqref{eq:sigmaint} as
\begin{align}
\label{eq:sigma_calculation}
\sigma^2_n=&\;\frac{2\alpha W^2}{\mathcal{A}}\frac{1}{k_nW}\int_0^{k_nW} \mathrm{d}x
\, J_0(x)
\nonumber\\
&\times
\left[
\left(1-\frac{x}{k_nW}\right)\cos{\left(\frac{\pi x}{k_nW}\right)}
+\frac{1}{\pi}\sin{\left(\frac{\pi x}{k_nW}\right)}
\right]
\ .
\end{align}
In the first and third terms on the right-hand side of
Eq.~\eqref{eq:sigma_calculation} we use $k_nW \gg 1$ to extend the
upper integration limits for these highly oscillating integrands to
infinity. We use
\begin{equation}
\label{eq:integral1}
\int_0^\infty\mathrm{d}x\, J_0(x)\cos{\left(\frac{\pi x}{k_nW}\right)}=\frac{1}{\sqrt{1-(\pi/k_nW)^2}}\ ,
\end{equation}
\begin{equation}
\label{eq:integral2}
\int_0^\infty\mathrm{d}x\, J_0(x) \sin{\left(\frac{\pi x}{k_nW}\right)}=0\ .
\end{equation}
The second term on the right-hand side of
Eq.\ \eqref{eq:sigma_calculation} can be integrated by parts. Using
that
\begin{equation}
\int \mathrm{d}x\,J_0(x)\,x= J_1(x)\,x\ , 
\end{equation}
with $J_1$ the first Bessel function of the first kind, we get
\begin{equation}
\label{eq:integral3}
\int_0^{k_nW}\mathrm{d}x\, J_0(x)\,x\cos{\left(\frac{\pi x}{k_nW}\right)}=
\frac{-J_1(k_nW)}{1-(\pi/k_nW)^2} \ .
\end{equation}
Combining Eqs.\ \eqref{eq:sigma_calculation}--\eqref{eq:integral2} and 
\eqref{eq:integral3}, we obtain
expression~\eqref{eq:sigma_sc} as the leading-order term in an
expansion in powers of $(k_n W)^{- 1}$.

In Eq.~\eqref{eq:prodgamint} we use $L \gg W$, the semiclassical limit
$k_n W^2 / L \gg 1$, and the asymptotic expansion of the Bessel
function to obtain the leading-order term
\begin{align}
\langle \gamma_n^{\mathrm{l}}\gamma_n^{\mathrm{r}} \rangle =&\; 
\frac{2\alpha W^2}{\mathcal{A}} \int_{0}^{1}\mathrm{d}z 
\,\sqrt{\frac{2}{\pi \ \theta(z)}}
\cos{\left(\theta(z)-\frac \pi 4\right)} 
\nonumber \\
&\times
\left[(1-z)\cos{\left(\pi z\right)} + \frac{1}{\pi}\sin{\left(\pi z\right)} \right] \ ,
\end{align}
where $\theta(z) = k_n L + (k_n W^2 / 2L) z^2$. Using the same
inequalities we evaluate the integral by the stationary-phase method.
The stationary point is at $\bar z=0$, and we have
\begin{align}
\langle \gamma_n^{\mathrm{l}}\gamma_n^{\mathrm{r}} \rangle \simeq&\; 
\frac{2\alpha W^2}{\mathcal{A}} \sqrt{\frac{2}{\pi k_nL}}\int_{0}^{\infty}\mathrm{d}z 
\cos{\left(\theta(z)-\frac \pi 4\right)} \ .
\end{align}
The evaluation of this Fresnel integral yields
Eq.~\eqref{eq:prodgam_sc}.

\bibliography{phase}

\begin{thebibliography}{44}
\bibitem [{foo()}]{footnote:shift}%
  \BibitemOpen
  \href@noop {} {\ }\bibinfo {note} {Each individual sequence for a given
  geometry exhibits the quasi-periodicity of $\pi/L$ in its $k_n$-dependence of
  the partial distribution of $\mathcal{P}(D_n < 0)$, but with an arbitrary
  phase-shift. In order to compare equivalent behaviors of $\mathcal{P}(D_n <
  0)$ for a given $k_n$, we adjusted a rigid shift for each geometry (for the
  whole $V_\mathrm{g}$ interval) by minimizing its departure from the analytic
  curve of the statistical model.}\BibitemShut {Stop}%
\end{thebibliography}

\end{document}